\documentclass[preprint,pteplogo]{ptephy_v2}

\preprintnumber{XXXX-XXXX} 
\usepackage[colorlinks=true, linkcolor=blue, citecolor=blue, urlcolor=blue]{hyperref}
\usepackage{orcidlink}
\usepackage{url}
\usepackage{amsmath}
\usepackage{graphicx}
\usepackage{multirow}
\usepackage{booktabs}
\usepackage{colortbl}
\usepackage{placeins}
\usepackage{float}

\begin{document}

\title{Development of a Neural Network-Based Background Suppression Technique for $\Sigma N$ Cusp Spectroscopy at J-PARC}

\author[1]{K.~Amemiya \orcidlink{0009-0004-9977-7799}}
\author[1,2,*]{Y.~Ichikawa \orcidlink{0000-0001-7754-6368}}
\author[1]{S.~H.~Hayakawa \orcidlink{0000-0002-8301-8067}}
\author[2]{K.~Tanida \orcidlink{0000-0002-8255-3746}}
\author[3]{S.~H.~Kim \orcidlink{0000-0003-2673-9818}}
\author[2,1]{F.~Oura \orcidlink{0009-0005-9504-7120}}
\author[4]{H.~I.~Lee \orcidlink{0009-0001-8997-9045}}
\author[1]{R.~J.~Saito \orcidlink{0009-0006-3279-1902}}
\author[1]{K.~Shimazaki}
\author[1]{R.~Sasaki}
\author[1]{Y.~Nakayama}

\affil[1]{Department of Physics, Tohoku University, Sendai 980-8578, Japan}
\affil[2]{Advanced Science Research Center, Japan Atomic Energy Agency, Tokai, Ibaraki 319-1195, Japan}
\affil[3]{Department of Physics, Kyungpook National University, Daegu 41566, Republic of Korea}
\affil[4]{Department of Physics, Korea University, Seoul 02841, Republic of Korea}
\affil[*]{E-mail: \href{mailto:yudai.ichikawa.d3@tohoku.ac.jp}{yudai.ichikawa.d3@tohoku.ac.jp}}


\begin{abstract}

A clear spectral enhancement, known as the ``$\Sigma N$ cusp'', has been observed near the $\Sigma N$ threshold in the $d(K^-, \pi^-)$ reaction.
To understand the dynamical origin of this enhancement, the J-PARC E90 experiment aims to investigate the missing-mass spectrum with an unprecedented resolution of 0.4 MeV ($\sigma$).
In this experiment, a Hyperon Time Projection Chamber (HypTPC) is utilized to detect charged decay products and suppress severe contamination from quasi-free (QF) background processes.
While a conventional track multiplicity condition of three (Mt=3) effectively suppresses these QF events, it restricts the signal statistics to approximately 17\% and introduces a mass-dependent acceptance bias that distorts the spectrum.
In contrast, events with a track multiplicity of two (Mt=2) offer roughly double the statistical power ($\sim$39\%) with minimal mass dependence, but they suffer from heavy background contamination.
To fully exploit the Mt=2 events, we developed an innovative background suppression technique based on a neural network.
By constructing a binary classification model using the HypTPC track topology and energy loss ($dE/dx$) as input features, we successfully discriminated the signal from QF backgrounds.
This machine learning approach achieves a signal-to-noise ratio comparable to the strict Mt=3 condition while preserving the integrity of the spectral shape.
By combining this independent ML-selected Mt=2 sample with the conventional Mt=3 sample, the total usable statistics are effectively doubled compared to traditional methods, significantly enhancing the sensitivity for determining the $\Sigma N$ cusp parameters.

\end{abstract}

\subjectindex{H40, D32}
\maketitle

\section{Introduction}\label{sec:introduction}

A clear spectral enhancement, known as the ``$\Sigma N$ cusp'', has been observed near the $\Sigma N$ threshold ($\sim 2.13$ GeV/$c^2$) in reactions such as $d(K^-, \pi^-)\Lambda p$.
The exact spectral shape of this cusp is highly sensitive to the $\Sigma N$ scattering length.
To understand the dynamical origin of this enhancement, the J-PARC E90 experiment is planned to perform high-resolution missing-mass spectroscopy of this structure using an incident $K^-$ momentum of 1.4 GeV/$c$.
The primary objective is to definitively extract the scattering length by analyzing the fine structure of the cusp.
Achieving this requires not only unprecedented mass resolution but also high statistics and a robust background suppression technique.

\begin{figure}[htbp]
  \centering
  \includegraphics[width=0.6\linewidth]{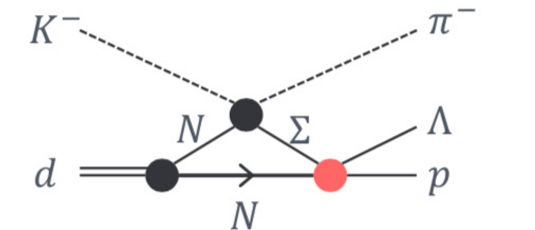}
  \caption{Reaction diagram of the $\Sigma N$ cusp production in the $K^- d \to \pi^- \Lambda p$ reaction.}
  \label{fig:cusp_diagram}
\end{figure}

According to the framework established by Dalitz and Deloff \cite{dalitz1981, dalitz1982}, the $\Sigma N$ cusp in the $d(K^-, \pi^-)$ reaction is generated through a two-step process, as illustrated in the reaction diagram in Fig.~\ref{fig:cusp_diagram}:
\begin{equation}
    K^- + d \to \pi^- + (\Sigma N) \to \pi^- + (\Lambda p).
\end{equation}

A unique advantage of the $(K^-, \pi^-)$ reaction at forward angles ($\cos \theta_{CM} > 0.95$) is that the elementary process is dominated by the non-spin-flip amplitude.
Consequently, the spin-triplet state ($S=1$) of the target deuteron is preserved in the intermediate $YN$ system.
Furthermore, since the final state $\Lambda p$ has isospin $T=1/2$, the final-state interaction (FSI) selectively probes the $(T, S) = (1/2, 1)$ channel.
Here, the observable reaction rate $R_S^t$ can be expressed using the scattering length $A_0 = a + ib$ as:
\begin{equation}
    R_S^t \propto \left|
    \frac{1}{1 - i k_\Sigma (a+ib)} \right|^2 = \frac{1}{|(1 + k_\Sigma b) - i k_\Sigma a|^2},
\end{equation}
where $k_\Sigma$ is the relative momentum of the $\Sigma N$ system.
Here, the real part $a$ represents the strength of the $\Sigma N$ two-body interaction, while the imaginary part $b$ represents the $\Sigma N \to \Lambda N$ conversion strength.

\begin{figure}[htbp]
  \centering
  \includegraphics[width=0.8\linewidth]{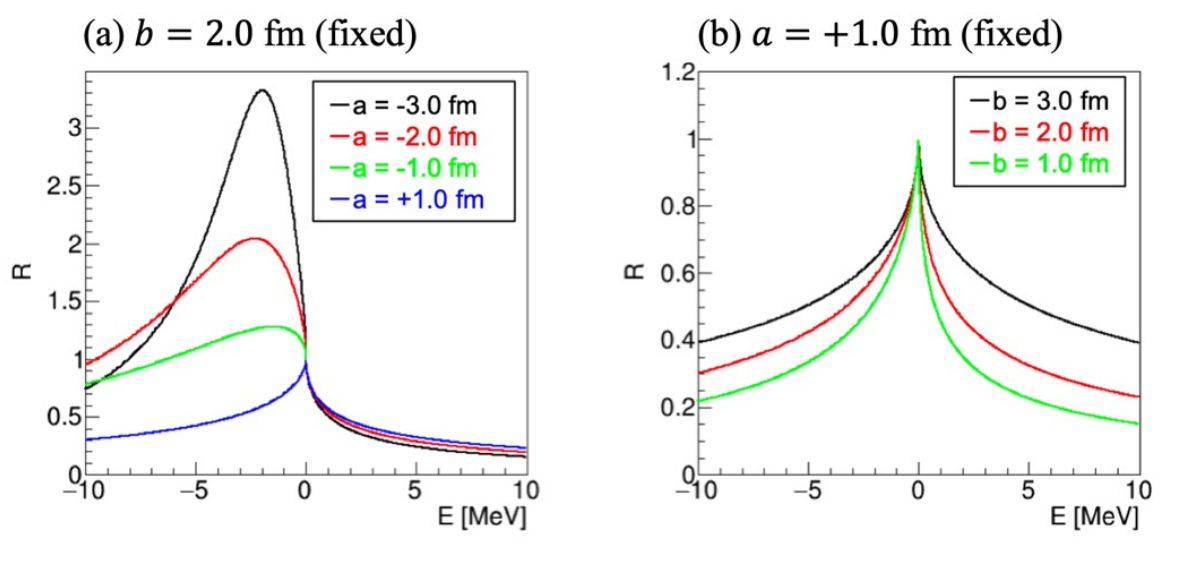}
  \caption{Dependence of the theoretical reaction rate $R_S^t$ on the scattering length parameters $a$ and $b$. The spectral shape changes significantly, indicating either a sharp cusp or a broad bound-state structure.}
  \label{fig:cusp_param}
\end{figure}

Figure~\ref{fig:cusp_param} shows the behavior of $R_S^t$ as a function of the excitation energy $E$ from the $\Sigma N$ threshold.
Above the threshold ($E > 0$), $R_S^t$ is primarily governed by the imaginary part $b$.
Below the threshold ($E < 0$), the formula resembles a Breit-Wigner distribution.
If the interaction is strongly attractive ($a < 0$ with a large magnitude), a peak structure corresponding to an unstable bound state appears below the threshold.
Conversely, a positive or small $a$ results in a sharp kinematic cusp at the threshold.
Since the observed spectral shape is extremely sensitive to the scattering length parameters, performing high-precision spectroscopy and fitting the measured spectrum enables the definitive determination of $a$ and $b$.

While the $\Sigma N$ cusp has been studied over 50 years, the dynamical origin of the enhancement---whether it is an inelastic virtual state or an unstable bound state---remains unresolved.
Previous experiments \cite{tan1969, braun1977, eastwood1971, pigot1985, budzanowski2010, acharya2022, ichikawa2019} have suffered from either low statistics, insufficient mass resolution, or severe background contamination, leaving the dynamical origin of the cusp unclear.
Detailed reviews of these historical experiments and theoretical formulations can be found in our comprehensive reports \cite{e90_prop, amemiya_mthesis}.

\section{Experimental Setup and HypTPC Detector}\label{sec:setup}

\begin{figure}[htbp]
  \centering
  \includegraphics[width=0.7\linewidth]{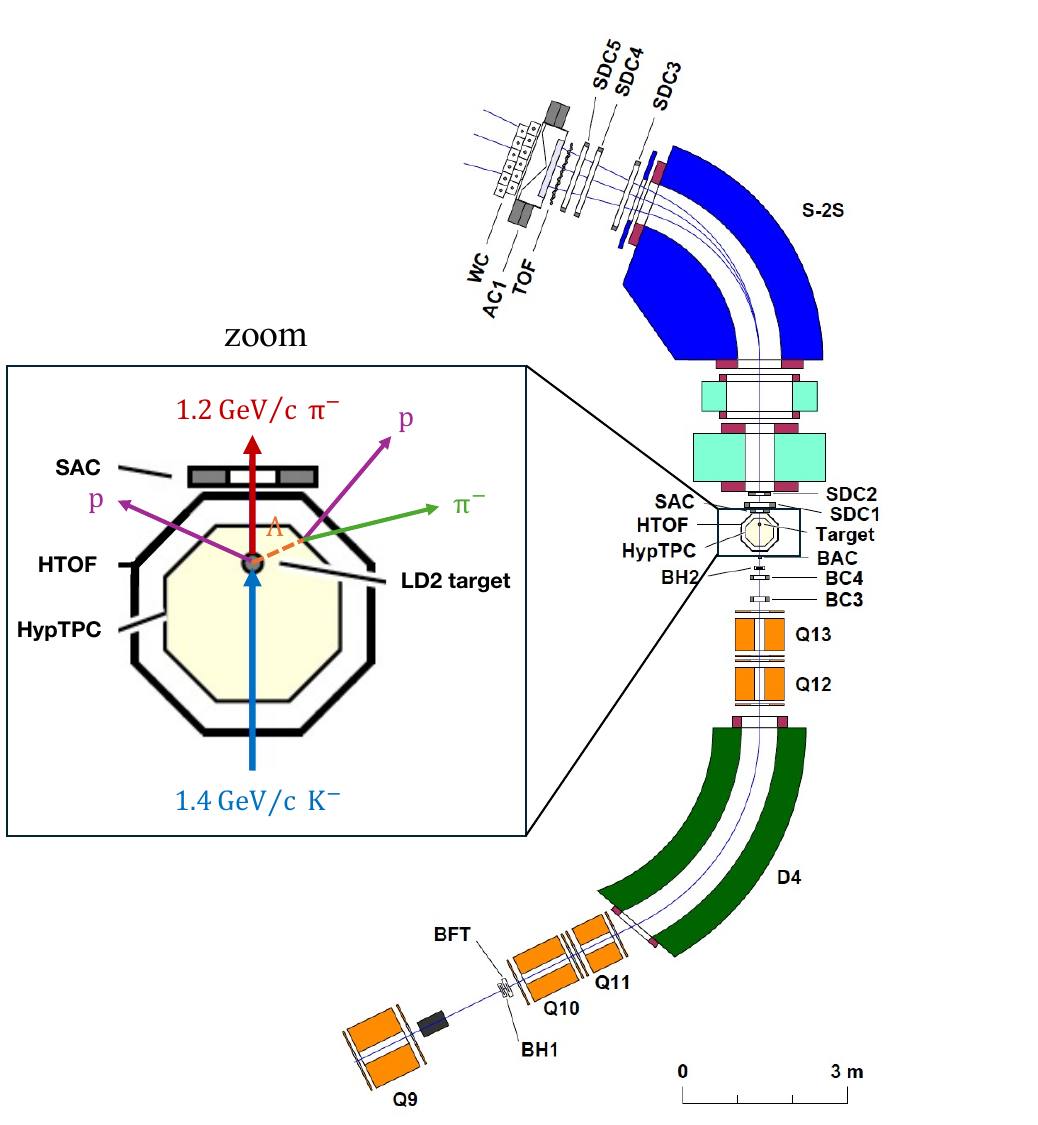}
  \caption{Overview of the J-PARC E90 experimental setup.}
  \label{fig:setup}
\end{figure}

To overcome the limitations in previous measurements, the E90 experiment utilizes the high-intensity $K^-$ beam at the J-PARC K1.8 beamline and the S-2S spectrometer to momentum-analyze the scattered $\pi^-$.
This combination achieves a world-leading missing-mass resolution of 0.4 MeV ($\sigma$), which is essential to accurately determine the scattering length through precise spectrum fitting \cite{e90_prop}.
An overview of the J-PARC E90 experimental setup is shown in Fig.~\ref{fig:setup}.

The $(T, S) = (1/2, {}^3S_1)$ channel in the $\Sigma N$ system is selected by combining a dedicated trigger logic with decay-topology tracking.
First, the forward-scattered $\pi^-$ is selected by the downstream Scattering Aerogel Cherenkov counter (SAC) in combination with the S-2S spectrometer.
Selecting these forward scattering kinematics ($\cos\theta_{CM} > 0.95$) ensures the dominance of the non-spin-flip amplitude, thereby inheriting the spin-triplet state ($S=1$) of the deuteron target.

Following the trigger selection, the Hyperon Time Projection Chamber (HypTPC) tracks the decay products around the target region.
The HypTPC is installed at the target position, enclosing a liquid deuterium (LD$_2$) target.
This LD$_2$ target ($\rho = 0.169$~g/cm$^3$) is contained in a cell with a diameter of 54~mm and a length of 100~mm.
The target cell is coaxially surrounded by multiple layers: a 0.5-mm-thick Mylar layer, a 0.5-mm-thick Carbon Fiber Reinforced Polymer (CFRP, $\rho = 1.70$~g/cm$^3$) vacuum pipe, and a 3-mm-thick G10 ($\rho = 1.70$~g/cm$^3$) target holder.
A beam window is provided in the beam passage region by removing the G10 material.
Detailed knowledge of these material budgets is critical for evaluating the energy loss of particles.
As will be discussed in Section~\ref{sec:multiplicity}, this energy-loss mechanism determines whether low-momentum spectator protons can reach the active volume of the HypTPC, and thus plays an important role in the multiplicity-based background suppression strategy.

\begin{figure}[htbp]
  \centering
  \includegraphics[width=0.9\linewidth]{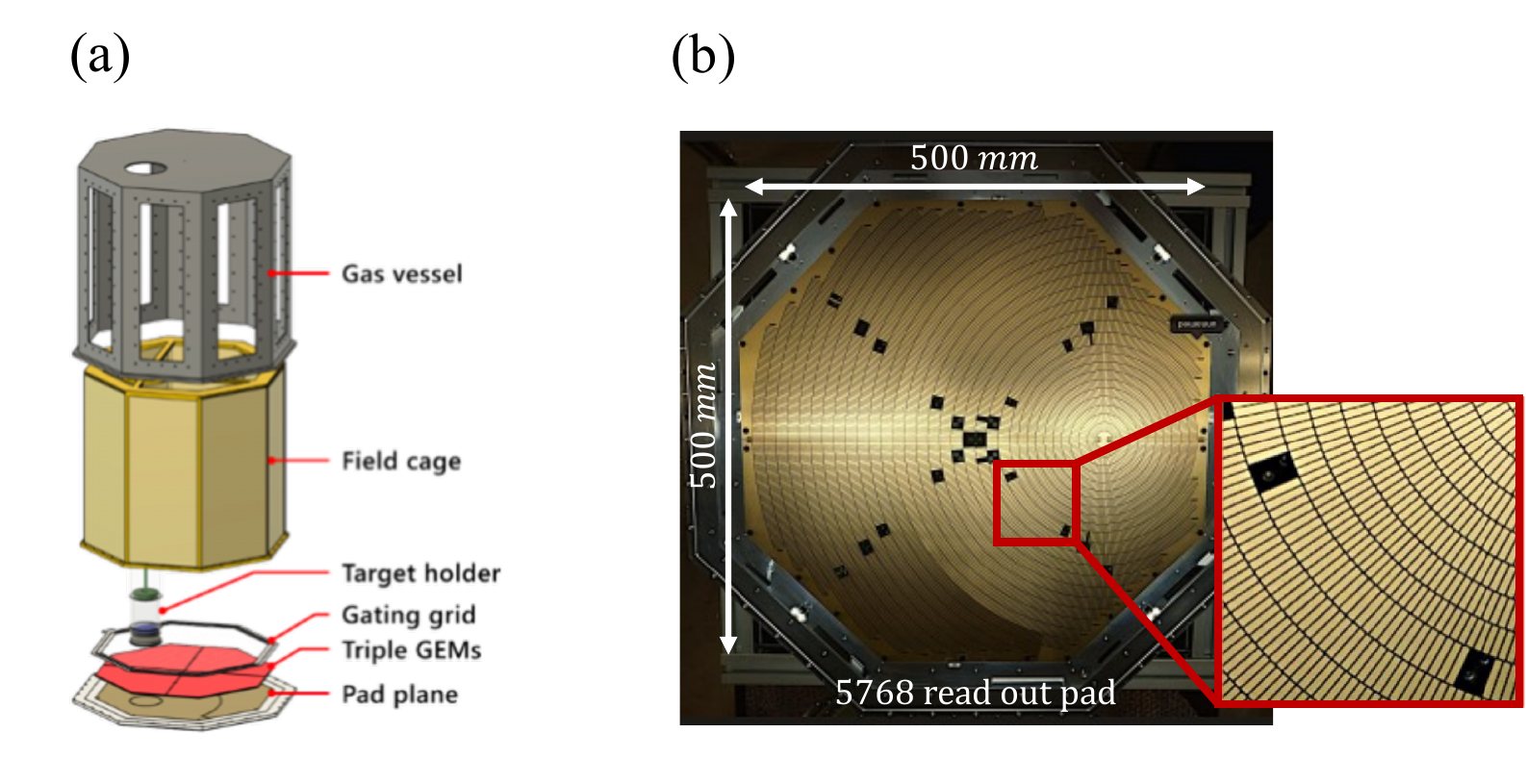}
  \caption{(a) Expanded view of the HypTPC. (b) The readout pad plane placed at the bottom part of the HypTPC.}
  \label{fig:tpc}
\end{figure}

To operate the HypTPC under high particle-rate conditions, P-10 (Ar-CH$_4$ (90:10)) gas is adopted for its high-rate capability, along with a gating-wire grid to suppress ion back-flow.
The ionized electrons drift under a uniform electric field, and the signals are amplified by a triple Gas Electron Multiplier (GEM) stack before being read out by a concentric anode-pad plane.
As shown in Fig.~\ref{fig:tpc}, the readout pad plane consists of two sectors: the inner sector has 10 layers with a pad size of 9~mm in length and 2.1--2.7~mm in width, and the outer sector has 22 layers with a pad size of 12.5~mm in length and 2.3--2.4~mm in width.
Full design details on the HypTPC can be found in Refs.~\cite{kim2019, ichikawa_hyptpc}.

By reconstructing the three-dimensional trajectories and the specific energy loss ($dE/dx$) measured by the HypTPC, we can reconstruct the decay topology ($\Lambda \to p\pi^-$ and a recoil proton).
This allows us to identify the $\Lambda p$ final state, which effectively isolates the isospin $T=1/2$ state.
Furthermore, this track topology information is vital for suppressing the dominant quasi-free hyperon-production backgrounds, as will be detailed in Section~\ref{sec:multiplicity}.

To fulfill the specific requirements of the E90 experiment, the HypTPC is operated in a specialized configuration.
In conventional experiments, the chamber is installed inside a 1~T superconducting magnet.
In E90, however, the target must be positioned as close as possible to the first tracking detector of the S-2S spectrometer (SDC1) in order to maximize the S-2S acceptance.
Consequently, the HypTPC is operated without the magnet owing to the limited space around the target region, and is installed in a reversed configuration, rotated by 180~degrees around the beam axis so that the target-mount section is placed on the downstream side.
In this field-free configuration, the spatial resolution of the HypTPC degrades to approximately $0.6$--$1.0$~mm (compared to $0.2$--$0.3$~mm under a typical 1~T magnetic field) due to the enhanced transverse diffusion of drifting electrons.
The transverse diffusion coefficient without the magnetic field is estimated to be $D_T \sim 0.57$~mm/$\sqrt{\mathrm{cm}}$ \cite{amemiya_mthesis}.

The most critical limitation arising from the absence of the magnetic field is that particle momenta cannot be measured from track curvature.
This makes conventional particle identification (PID) based on momentum and specific energy loss unavailable.
Therefore, event selection must rely only on the available tracking topology and energy-deposit information, which motivates the development of a dedicated background-suppression method.

\section{Conventional Multiplicity-Based Selection and Its Limitations}\label{sec:multiplicity}
The subsequent offline analysis aims to isolate the $\Sigma N$ cusp signal from the acquired data.
However, the lack of a magnetic field severely limits the available track information from the HypTPC, necessitating a highly optimized background suppression strategy.

In this study, we consider quasi-free (QF) hyperon production processes as the dominant sources of background.
It should be noted that there might also be unknown or irreducible physical backgrounds that do not originate from the $\Sigma N$ cusp.
For instance, complex two-step processes---such as small-angle elastic scattering or a charge-exchange reaction ($N K^- \to p K$) followed by a $(K, \pi^-)$ reaction---could theoretically contribute to the continuous background spectrum.
However, previous measurements by Eastwood et al.~\cite{eastwood1971} under similar incident kinematics (1.45 and 1.65 GeV/$c$) demonstrated that simple kinematic selections---specifically a forward scattering angle ($\cos\theta_{K\pi}^{\mathrm{CM}} > 0.9$) and a proton momentum cut ($p_p > 170$ MeV/$c$)---are sufficient to obtain a largely background-free spectrum around the cusp region.
Based on this empirical evidence, the contribution of non-QF backgrounds is expected to be minimal.
Therefore, this methodology paper focuses exclusively on suppressing the dominant QF backgrounds.

\subsection{Principle of Multiplicity-Based Selection}

\begin{figure}[htbp]
  \centering
  \includegraphics[width=1.0\linewidth]{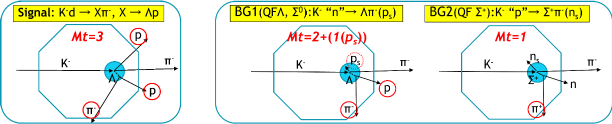}
  \caption{Conceptual diagram of background suppression using track multiplicity in the HypTPC. Left: The $\Sigma N$ cusp signal is observed as three tracks ($\mathrm{Mt}=3$). Center/Right: QF backgrounds are observed with fewer tracks ($\mathrm{Mt}=1$ or $\mathrm{Mt}=2$) because spectator nucleons are either undetected or neutral.}
  \label{fig:multiplicity_logic}
\end{figure}

Background suppression in the E90 experiment relies on the track multiplicity (Mt) detected by the HypTPC.
The principle of this selection is based on the difference in decay topologies and kinematics between the signal and the QF background processes, as illustrated in Fig.~\ref{fig:multiplicity_logic}.

The signal process, $K^- + d \to X + \pi^-$ followed by $X \to \Lambda p$ and $\Lambda \to p \pi^-$ (where $X$ represents the $\Sigma N$ cusp), produces three charged particles ($p, p, \pi^-$) in the final state.
Since the primary scattered $\pi^-$ is identified by the forward S-2S spectrometer, the HypTPC handles the three decay products.
These particles typically possess sufficient momentum to escape the LD$_2$ target and its surrounding materials, resulting in a multiplicity of Mt=3.

In contrast, background events are categorized into two processes:
\begin{itemize}
    \item \textbf{BG1 (QF-$\Lambda, \Sigma^0$):} In these reactions, the target proton within the deuteron acts as a spectator ($p_s$). Although three charged particles ($p, \pi^-, p_s$) are physically present in the final state, the spectator proton typically carries a very low momentum ($\lesssim 200$~MeV/$c$) originating from the Fermi motion within the deuteron. Such low-momentum protons have a short range and are likely to stop within the LD$_2$ target or the G10 target holder. Consequently, only the two decay products from the $\Lambda$ are detected, leading to Mt=2.
    \item \textbf{BG2 (QF-$\Sigma^+$):} Here, a neutron acts as a spectator ($n_s$), which is invisible to the HypTPC. The $\Sigma^+$ decay produces only one charged particle ($\pi^+$ or $p$), resulting in Mt=1.
\end{itemize}

By requiring the Mt=3 condition, the dominant QF backgrounds can be theoretically eliminated while preserving the signal.

\subsection{Monte Carlo Simulation}
\label{subsec:simulation}

To assess the effectiveness of the background suppression based on the track multiplicity, a detailed Monte Carlo simulation using the Geant4 toolkit was performed.
Since the background suppression principle relies on the assumption that low-momentum spectator protons from QF reactions lose their energy and stop within the target materials before reaching the sensitive volume of the HypTPC, precise implementation of the detector geometry and material budget is essential.

\begin{figure}[htbp]
  \centering
  \includegraphics[width=0.8\linewidth]{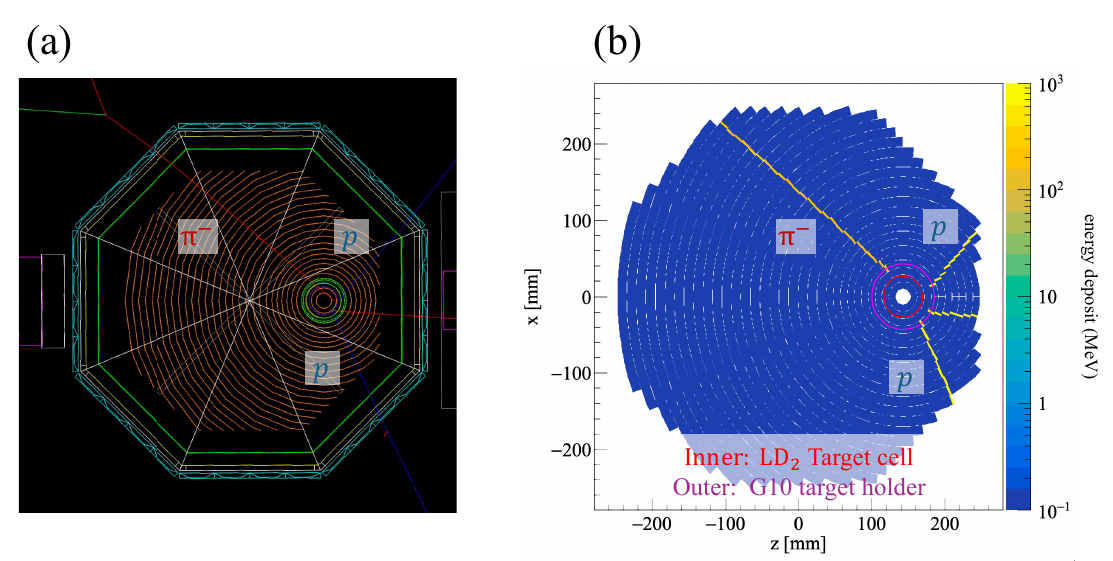}
  \caption{(a) Simulated tracks of a $\Sigma N$ cusp event inside the HypTPC. (b) Example of the simulated TPC pad hit distribution.}
  \label{fig:e90_g4tpc}
\end{figure}

In the simulation, the S-2S spectrometer was implemented with realistic magnetic field maps.
For the target region, the geometry of the HypTPC readout pad plane and the material distribution of the LD$_2$ target system---including the cell windows and holder---were faithfully reproduced, as shown in Fig.~\ref{fig:e90_g4tpc}.
This detailed modeling enables an accurate evaluation of the energy loss for both decay products and spectator protons.
Note that the intrinsic spatial resolution of the HypTPC tracking was not included in this simulation; the tracking information was treated as idealized to focus on the fundamental kinematic and topological features relevant to establishing the methodology.

Signal events for $\Sigma N$ cusp were generated according to a missing-mass distribution featuring a cusp structure.
The final production probability was assumed to be proportional to $|F_d(q)|^2 |f_{\Sigma N \to \Lambda N}(k)|^2$, where $F_d(q)$ is the deuteron form factor derived from its wave function, and $f_{\Sigma N \to \Lambda N}(k)$ is the $\Sigma N \to \Lambda N$ transition amplitude based on zero-range approximation.
For the scattering length $a_s$, we adopted a typical theoretical value of $a_s = 2.06 + i\,4.64$~fm, based on the Nijmegen model-D (ND) from Ref.~\cite{miyagawa1999}.

The dominant QF background was generated using the impulse approximation, where the Fermi momentum of the nucleon in deuteron was sampled using the Bonn potential wave function.
The generated events were then filtered by applying the experimental trigger conditions and the geometrical acceptance of the S-2S spectrometer\footnote{The source code for this Geant4 simulation is available at \url{https://github.com/jparc-k18/k18-geant4/tree/s2s}.}.

In the simulation, the HypTPC track multiplicity for each event was determined by considering the track-reconstruction conditions used in the actual data analysis.
Specifically, the following criteria were applied:
\begin{enumerate}
    \item \textbf{Hit requirement}: A charged particle was counted as a HypTPC track only when it produced at least four pad hits. This requirement reflects the minimum number of space points needed for the linear tracking algorithm.
    \item \textbf{Particle species}: Only protons and charged pions ($\pi^\pm$) were included in the multiplicity count. In the actual experiment, light charged particles such as electrons and positrons ($e^\pm$) can be rejected using the $dE/dx$ information. Therefore, in the simulation, true Monte Carlo particle information was used to exclude such particles.
    \item \textbf{Exclusion of primary beam and spectrometer tracks}: The incident $K^-$ beam and the primary forward-scattered $\pi^-$ analyzed by the S-2S spectrometer were excluded from the HypTPC multiplicity count, since they do not constitute the decay topology measured by the HypTPC.
\end{enumerate}
The total number of tracks satisfying these criteria was defined as the HypTPC track multiplicity for each event.
Further details on the simulation framework are provided in Ref.~\cite{amemiya_mthesis}.

\subsection{Performance and Limitations of the Multiplicity Cut}

\begin{figure}[htbp]
  \centering
  \includegraphics[width=1.0\linewidth]{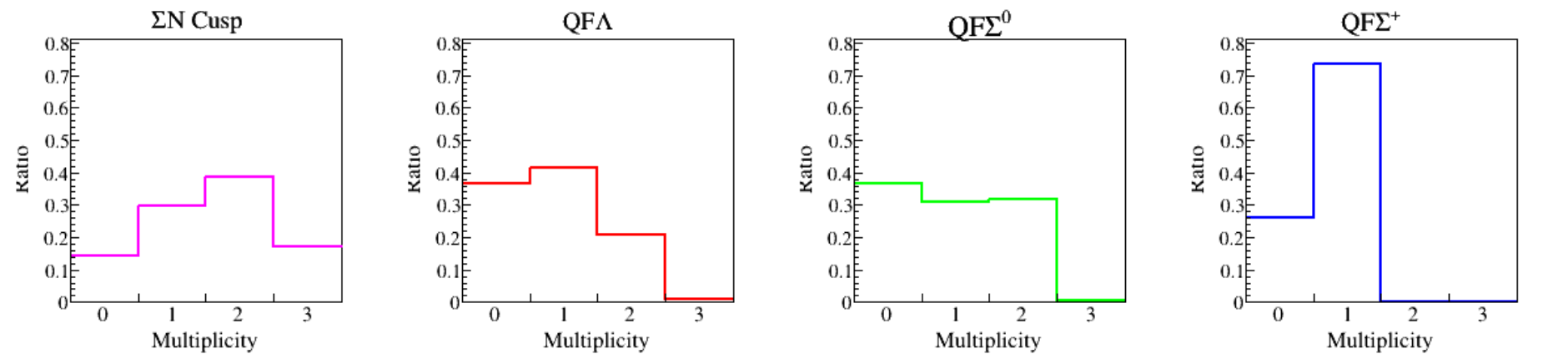}
  \caption{Track multiplicity distribution obtained from Geant4 simulations for the signal and background channels.}
  \label{fig:G4multiplicity}
\end{figure}

The resulting multiplicity distribution obtained from these simulations is shown in Fig.~\ref{fig:G4multiplicity}, and the ratios for each reaction channel are summarized in Table~\ref{tab:mt_ratios}.

\begin{table}[htbp]
  \centering
  \caption{Ratio of events satisfying $\mathrm{Mt}=3$ and $\mathrm{Mt}=2$ conditions for each reaction channel based on Geant4 simulations. Note that QF-$\Sigma^+$ is omitted here as it yields only one charged track ($\mathrm{Mt}=1$). Errors represent statistical uncertainties.}
  \label{tab:mt_ratios}
  \begin{tabular}{lcc}
    \toprule
    Channel & $\mathrm{Mt}=3$ Ratio [\%] & $\mathrm{Mt}=2$ Ratio [\%] \\
    \midrule
    $\Sigma N$ Cusp (signal) & $17.3 \pm 0.2$ & $38.6 \pm 0.2$ \\
    QF-$\Lambda$ & $0.89 \pm 0.02$ & $20.9 \pm 0.1$ \\
    QF-$\Sigma^0$ & $0.64 \pm 0.04$ & $31.8 \pm 0.3$ \\
    \bottomrule
  \end{tabular}
\end{table}

However, these simulations revealed two critical drawbacks of the Mt=3 condition.
First, it severely limits the available statistics, retaining only $\sim$17\% of the total signal events.
Second, it introduces a mass-dependent acceptance bias; the detection efficiency for the third track significantly drops in the low missing-mass region.
This bias artificially distorts the delicate spectral shape of the cusp, posing a significant risk of systematic errors when deducing the scattering length.

To examine the impact of the Mt selection on the observed missing-mass spectrum, we scaled the simulated events to the expected statistics for the total 15-day beam time.
The expected yield of the $\Sigma N$ cusp in the inclusive measurement is estimated to be $7.6 \times 10^4$ events, based on the experimental design parameters \cite{e90_prop}.
The mass spectra were smeared with the design mass resolution of 0.4 MeV ($\sigma$) and binned at 0.2 MeV/bin to examine the fine structure of the cusp.

\begin{figure}[htbp]
  \centering
  \includegraphics[width=0.95\linewidth]{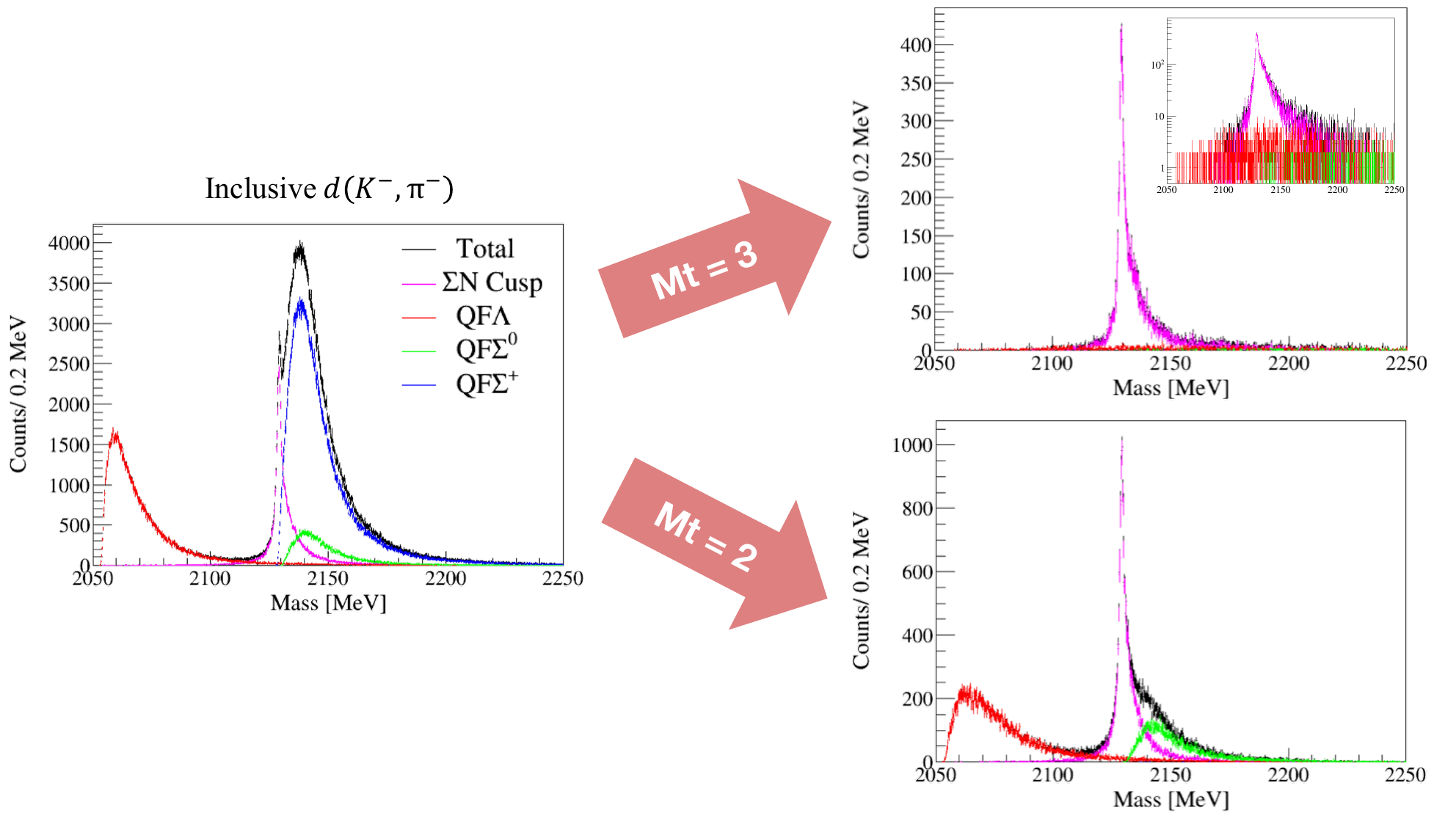}
  \caption{Comparison of the simulated missing-mass spectra for inclusive, $\mathrm{Mt}=3$, and $\mathrm{Mt}=2$ selections. The $\mathrm{Mt}=3$ condition provides a clean signal, whereas $\mathrm{Mt}=2$ retains a considerable amount of background.}
  \label{fig:cuspcomp_mt2_mt3}
\end{figure}

Figure~\ref{fig:cuspcomp_mt2_mt3} compares the missing-mass spectra before and after applying the Mt=3 and Mt=2 selections.
As visually evident, the Mt=3 condition strongly suppresses the QF background and cleanly extracts the signal component.
In contrast, while the Mt=2 selection retains approximately double the signal statistics of Mt=3, it suffers from significant QF background contamination, which obscures the spectral shape in the cusp region.

\begin{figure}[htbp]
  \centering
  \includegraphics[width=0.9\linewidth]{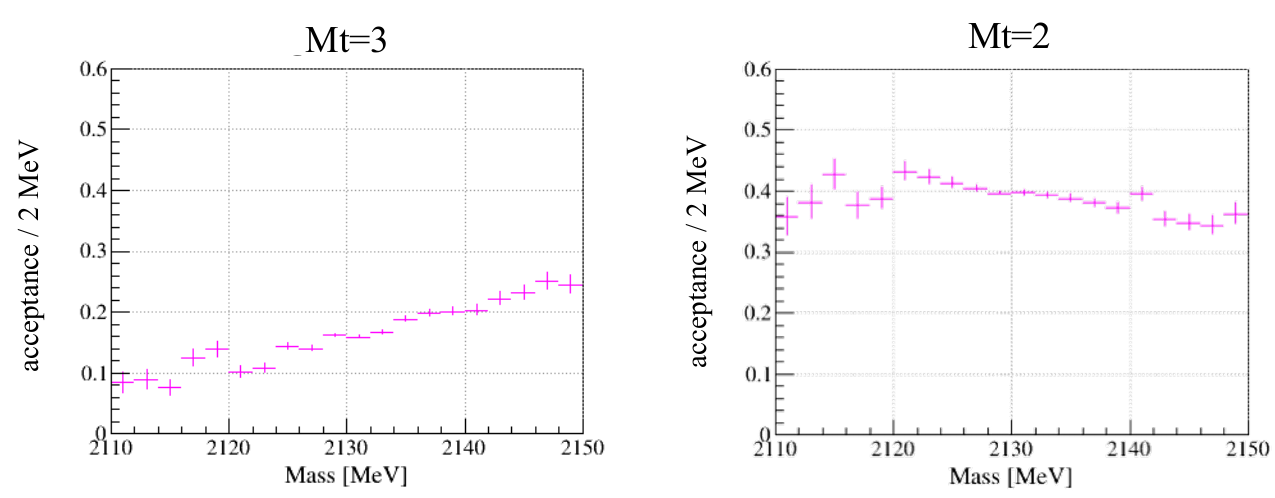}
  \caption{Mass dependence of the acceptance for the $\Sigma N$ cusp signal under $\mathrm{Mt}=3$ and $\mathrm{Mt}=2$ conditions. The $\mathrm{Mt}=3$ acceptance drops significantly in the low-mass region, whereas the $\mathrm{Mt}=2$ acceptance remains relatively flat.}
  \label{fig:acceptance_mt2_mt3}
\end{figure}

A critical issue in this analysis is whether the event selection distorts the intrinsic cusp shape.
Figure~\ref{fig:acceptance_mt2_mt3} shows the mass dependence of the acceptance (defined as the ratio of selected to generated events) for the signal process under both Mt conditions.

For the Mt=3 selection, a clear acceptance bias is observed: the detection efficiency decreases in the lower missing-mass region.
This can be understood kinematically.
A larger missing mass corresponds to a larger $Q$-value, which imparts higher kinetic energy to the decay protons.
Protons with higher momenta are more likely to penetrate the target container and surrounding materials to reach the active volume of the HypTPC, increasing the probability of detecting all three tracks.
Conversely, the Mt=2 selection exhibits a relatively flat acceptance across the threshold region, mitigating this spectral distortion.

To quantitatively assess the impact of these shape distortions and background contaminations on the extraction of physical parameters, we performed a simplified spectral fitting in the critical cusp region ($2124$--$2139$ MeV/$c^2$).
In this test, we fitted the simulated mass spectra directly with the theoretical cusp function.
Crucially, we intentionally performed this fit without applying any acceptance correction for the HypTPC and assumed a background-free spectrum.
The scattering length parameters ($a$ and $b$) were fixed to the exact input values used in the event generation, leaving only an overall normalization factor as a free parameter.
Under these constrained conditions, the reduced chi-square ($\chi^2/ndf$) serves as a direct indicator of how much the intrinsic cusp shape is deformed by the instrumental acceptance bias or obscured by residual background contamination.

The fitting was repeated over three independent smeared datasets to account for statistical fluctuations.
Table~\ref{tab:mt_fit_result} summarizes the background contamination and the $\chi^2/ndf$ of the fit.

\begin{table}[htbp]
  \centering
  \caption{Comparison of signal statistics, background contamination, and fitting accuracy ($\chi^2/ndf$) in the cusp region ($2124$--$2139$ MeV/$c^2$) for different selection conditions. Values are presented as the mean $\pm$ standard deviation from three independent fitting iterations.}
  \label{tab:mt_fit_result}
  \begin{tabular}{lccc}
    \toprule
    Condition & \begin{tabular}[c]{@{}c@{}}Stats.\\(2050--2250 MeV/$c^2$)\end{tabular} & \begin{tabular}[c]{@{}c@{}}BG Contamination\\(2124--2139 MeV/$c^2$)\end{tabular} & \begin{tabular}[c]{@{}c@{}}$\chi^2 / \text{ndf}$\\(2124--2139 MeV/$c^2$)\end{tabular} \\
    \midrule
    Inclusive & 76,000 & $68.22 \pm 0.01$\% & --- \\
    $\mathrm{Mt}=3$ & 13,180 & $1.77 \pm 0.01$\% & $2.23 \pm 0.19$ \\
    $\mathrm{Mt}=2$ & 29,369 & $10.94 \pm 0.03$\% & $8.93 \pm 0.09$ \\
    \bottomrule
  \end{tabular}
\end{table}

While the inclusive measurement suffers from a massive background contamination of $\sim$68\%, the Mt=3 selection successfully suppresses this to $\sim$1.8\%, yielding a reasonable fit quality ($\chi^2/ndf = 2.23$).
The slight deviation from an ideal $\chi^2/ndf \sim 1$ reflects the minor shape distortion caused by the acceptance bias.
However, the Mt=2 selection retains a relatively high background level of $\sim$10.9\%, which elevates the baseline of the spectrum.
As a result, the theoretical signal function alone fails to reproduce the data, leading to a poor $\chi^2/ndf$ of 8.93.

Based on the simulation results discussed above, the fundamental trade-off in the traditional multiplicity-based selection can be summarized as follows:
\begin{itemize}
    \item \textbf{Mt=3 selection:} Achieves strong background suppression (background contamination $\approx 1.8\%$) and a high S/N ratio. However, the available statistics are severely limited to approximately 17\% of the total signal, and the spectrum suffers from a mass-dependent acceptance bias.
    \item \textbf{Mt=2 selection:} Offers more than double the statistical power compared to Mt=3 and exhibits a relatively flat acceptance with minimal mass dependence. However, the high background contamination ($\approx 11\%$) significantly degrades the fitting accuracy.
\end{itemize}

\subsection{Kinematic Features and the Need for Machine Learning}

For the precise determination of the $\Sigma N$ scattering length, it is imperative to maximize the statistical yield while accurately understanding the intrinsic spectral shape.
Therefore, utilizing Mt=2 events, which excel in both statistical power and shape preservation, is the physically optimal strategy.
Nevertheless, the substantial background contamination in the Mt=2 sample requires a novel separation technique.

\begin{figure}[htbp]
  \centering
  \includegraphics[width=0.80\linewidth]{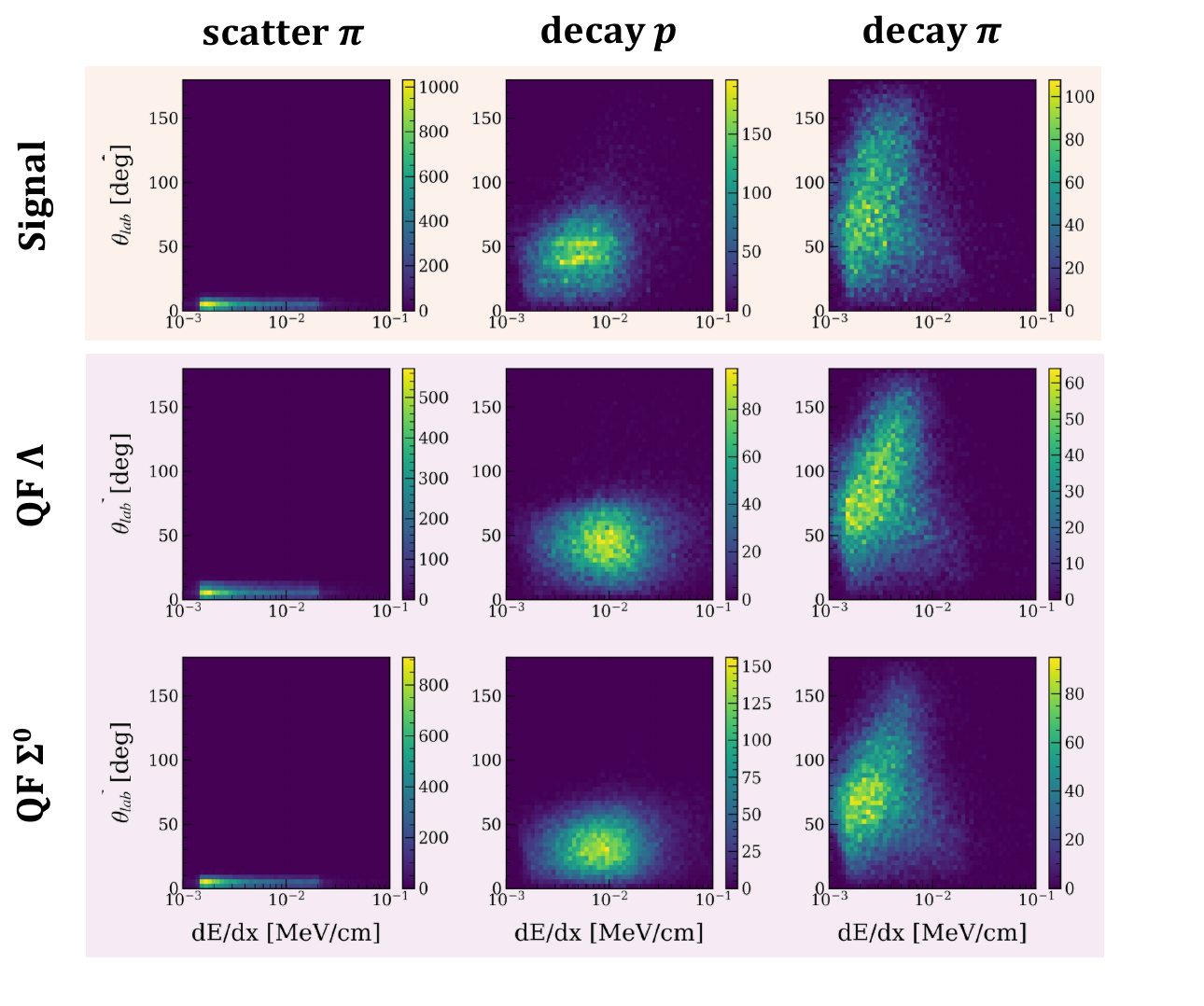}
  \caption{Two-dimensional correlation between the energy loss ($dE/dx$) and the scattering angle ($\theta$) for the $\Sigma N$ cusp signal and QF background events in the $\mathrm{Mt}=2$ sample.}
  \label{fig:mt2_dataplot}
\end{figure}

To establish such a technique, we investigated the detailed track information available from the HypTPC for Mt=2 events.
Key observable features include the direction vectors ($\vec{u}$) and the energy loss ($dE/dx$) of the reconstructed tracks.
Figure~\ref{fig:mt2_dataplot} shows the two-dimensional correlation between $dE/dx$ and the scattering angle $\theta$ for the signal and QF background events in the Mt=2 sample.

Taking the decay proton as an example, distinct kinematic trends can be observed.
In the signal ($\Sigma N$ cusp) process, the recoil nucleon receives very little momentum, often stopping within the target material.
Due to energy and momentum conservation, a relatively large momentum is distributed to the decay proton from the $\Lambda$, resulting in a tendency for smaller $dE/dx$ values.
Conversely, in QF background events, the scattered $\Lambda$ in the QF process is typically emitted forward with a large recoil momentum.
Consequently, its decay proton also concentrates in the forward direction in the laboratory frame, exhibiting a small scattering angle $\theta$.

Although qualitative differences exist between these distributions, as shown in Fig.~\ref{fig:mt2_dataplot}, the signal and background events overlap extensively in the parameter space.
Because of this large overlap, conventional selection methods---such as applying simple, one-dimensional threshold cuts on individual variables or employing likelihood methods that assume variables are independent---cannot effectively maximize statistics while maintaining high signal purity.

To overcome these limitations, we introduced a Machine Learning (ML) approach, specifically a Neural Network (NN), as a multivariate analysis tool.
NNs are capable of learning highly non-linear correlations among multiple observable variables, allowing them to construct complex decision boundaries in a high-dimensional feature space.
By employing an ML model, we aim to achieve an event classification that leverages multidimensional correlations, which is unattainable with traditional methods.
The detailed implementation and performance evaluation of our ML model are described in the following section.

\section{Neural Network-Based Event Classification}\label{sec:implementation}

To effectively classify the Mt=2 events into the $\Sigma N$ cusp signal and QF backgrounds, we constructed a binary classification model using a Multilayer Perceptron (MLP).
The model was implemented using the PyTorch framework, and the source code for this training is available in a public repository\footnote{The source code is available at \url{https://github.com/hyptpc/E90ML}.}.
The model was trained using simulated data generated by the Geant4 framework.
The QF-$\Sigma^+$ background was excluded from the training dataset due to its negligible contribution to the Mt=2 sample.

\subsection{Feature Selection and Data Preprocessing}
The neural network utilizes 12 input features derived from the reconstructed track topologies.
Although the HypTPC detects only two tracks in the Mt=2 sample, the full event topology includes the primary scattered $\pi^-$ analyzed by the forward S-2S spectrometer.
Therefore, we extracted the direction cosines ($u_x, u_y, u_z$) and the energy loss ($dE/dx$) for all three tracks associated with an event.

To ensure the stability of the neural network learning process, it is essential to assign physically consistent meanings to the input nodes.
Thus, we sorted the three tracks into Track 0, Track 1, and Track 2 based on the following rules:
\begin{itemize}
    \item \textbf{Track 0}: The scattered $\pi^-$ identified by the forward S-2S spectrometer, which is kinematically correlated with the HypTPC tracks via vertex matching.
    \item \textbf{Track 1 and Track 2}: The two candidate decay products detected in the HypTPC, sorted by their energy loss such that $(dE/dx)_{\text{Track 1}} < (dE/dx)_{\text{Track 2}}$.
\end{itemize}

The $dE/dx$ values were calculated using a truncated mean, discarding the top 30\% of the hit charges to mitigate the effect of the Landau tail.
Simulation studies confirmed that this $dE/dx$-based sorting strategy correctly identifies the decay proton as Track 2 with an accuracy exceeding 99.9\% across all relevant reaction channels.
Finally, all input variables were standardized to have a mean of zero and a variance of one (Z-score normalization) to stabilize the optimization process.

For training and evaluation, signal and background datasets, each containing on the order of $10^6$ simulated events, were prepared and randomly divided into training and validation datasets with a ratio of 80:20.
To address the extreme class imbalance (i.e., background events significantly outnumbering signal events), we applied class weighting to the Binary Cross-Entropy (BCE) loss function, penalizing the misclassification of signal events more heavily.

Furthermore, because simply using overall accuracy is often misleading for highly imbalanced datasets, we adopted Precision, Recall, and their harmonic mean, the F1 score, as our primary evaluation metrics.
These are defined as follows:
\begin{equation}
    \mathrm{Precision} = \frac{\mathrm{TP}}{\mathrm{TP} + \mathrm{FP}}, \quad
    \mathrm{Recall} = \frac{\mathrm{TP}}{\mathrm{TP} + \mathrm{FN}}, \quad
    \mathrm{F1} = 2  \times \frac{\mathrm{Precision} \times \mathrm{Recall}}
    {\mathrm{Precision} + \mathrm{Recall}} .
\end{equation}
\vspace{2pt}
where TP, FP, and FN represent the number of true positives, false positives, and false negatives, respectively.
Precision measures the purity of the selected signal, while Recall represents the signal detection efficiency.
Overfitting was prevented by monitoring this F1 score on the independent validation dataset and employing an early stopping mechanism.
This ensures that the model captures the optimal balance between signal efficiency and purity, which is critical for rare signal detection.

\subsection{Hyperparameter Optimization}
The performance of a neural network is sensitive to its hyperparameters, such as the number of layers, hidden units, learning rate, and dropout rate.
To optimize the network architecture efficiently, we employed the Optuna framework \cite{akiba2019}, which utilizes Bayesian optimization based on the Tree-structured Parzen Estimator (TPE) algorithm.
For computational efficiency, 1\% of the training dataset was randomly sampled for the optimization.
We conducted 100 trials, setting the objective function to maximize the F1 score evaluated on the validation set.
The defined search space and the resulting optimal hyperparameters are summarized in Table~\ref{tab:optuna_space}.

\begin{table}[htbp]
  \centering
  \caption{Search space and the best values for the hyperparameters optimized using Optuna.}
  \label{tab:optuna_space}
  \begin{tabular}{lll}
    \toprule
    Hyperparameter & Search space & Best value \\
    \midrule
    Batch Size & $[256, 512, 1024]$ & 256 \\
    Number of Layers & Integer $[3, 8]$ & 8 \\
    Hidden Units & Integer $[256, 1024]$ & 737 \\
    Dropout Rate & Float $[0.1, 0.5]$ & 0.16 \\
    Learning Rate & Log-uniform $[10^{-4}, 10^{-2}]$ & $4.6 \times 10^{-3}$ \\
    \bottomrule
  \end{tabular}
\end{table}

Based on this optimization, the final network architecture was configured as an 8-layer Multilayer Perceptron, with each hidden block incorporating Batch Normalization, a ReLU activation function, and Dropout.

\section{Results}\label{sec:results}

\subsection{Model Training and Threshold Optimization}
\begin{figure}[htbp]
  \centering
  \includegraphics[width=0.75\linewidth]{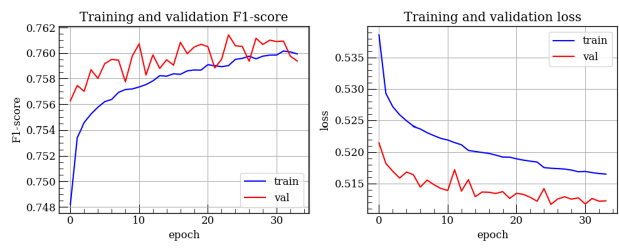}
  \caption{Evolution of the F1 score and Binary Cross-Entropy (BCE) loss during the training process. The early stopping mechanism prevented overfitting.}
  \label{fig:train_history}
\end{figure}

As shown in the training history (Fig.~\ref{fig:train_history}), the F1 score and the loss function evolved smoothly without significant divergence between the training and validation datasets, indicating no signs of overfitting.
The early stopping mechanism successfully halted the training at the optimal epoch.

\begin{figure}[htbp]
  \centering
  \includegraphics[width=0.8\linewidth]{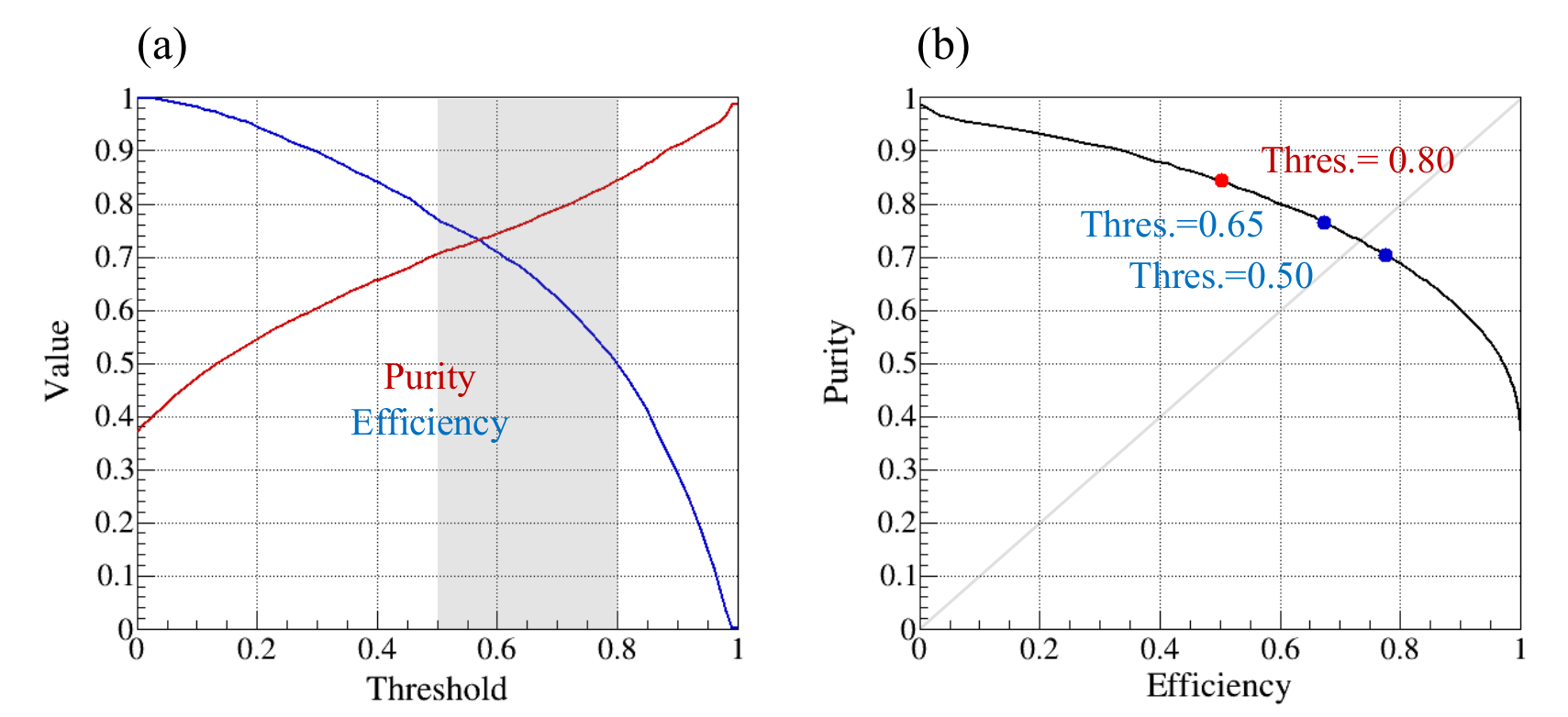}
  \caption{(a) Signal efficiency and purity as a function of the classification threshold. (b) The ROC-like curve showing the correlation between efficiency and purity. The selected threshold of 0.8 is marked in red.}
  \label{fig:roc}
\end{figure}

The final event selection was performed by applying a threshold to the network's output probability score.
As depicted in the Receiver Operating Characteristic (ROC) and efficiency-purity curves (Fig.~\ref{fig:roc}), there is an inherent trade-off between signal efficiency and purity.
In general machine learning applications, the optimal operating point is often chosen at the ``shoulder'' of the ROC curve (corresponding to thresholds between 0.5 and 0.65 in our case).
However, our primary goal is high-precision spectroscopy rather than simple classification.
Lower thresholds retain a non-negligible background contamination (3--4\%), which significantly degrades the spectral fitting accuracy ($\chi^2/ndf$).

Therefore, we prioritized high purity and selected a stricter threshold of 0.8.
Although this setting sacrifices some signal efficiency (retaining $\approx 50\%$, which yields $\approx 14,600$ events), it successfully suppresses the background contamination in the critical cusp region ($2124$--$2139$ MeV/$c^2$) to $\approx 2.1\%$.
This rivals the purity of the strict Mt=3 condition and ultimately provides the best fitting result.
Detailed comparisons of different threshold settings are discussed in Ref.~\cite{amemiya_mthesis}.

\subsection{Verification of the Missing-Mass Spectrum}
\begin{figure}[htbp]
  \centering
  \includegraphics[width=0.9\linewidth]{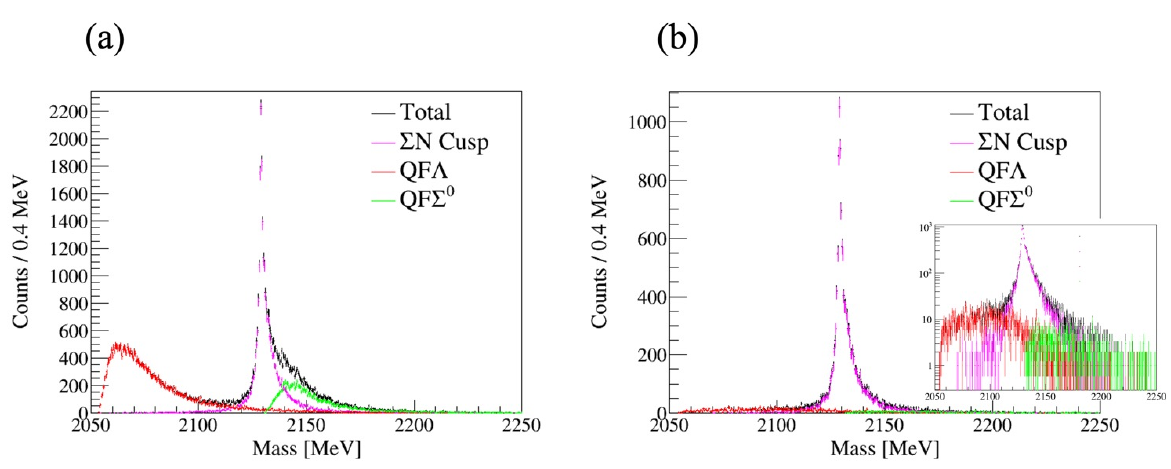}
  \caption{Comparison of the missing-mass spectra for (a) the simple $\mathrm{Mt}=2$ condition and (b) the $\mathrm{Mt}=2+\mathrm{ML}$ condition with a threshold of 0.8. The ML selection effectively removes the dominant background components.}
  \label{fig:ml_mmcomp}
\end{figure}

Applying the optimized ML selection to the actual simulated data resulted in the missing-mass spectrum shown in Fig.~\ref{fig:ml_mmcomp}.
The neural network dramatically suppressed the QF background processes, leaving a clear and distinct $\Sigma N$ cusp signal that was previously obscured in the simple Mt=2 spectrum.

To quantitatively assess the validity of this selection, we fitted the resulting spectra focusing on the cusp peak region ($2124$--$2139$ MeV/$c^2$).
It should be noted that this spectral fitting was intentionally performed directly on the simulated data without applying physical acceptance corrections, in order to evaluate how well the raw spectral shape is preserved under each selection method.
Therefore, the resulting reduced chi-square ($\chi^2/ndf$) values serve as auxiliary metrics for relative comparison among the selection strategies rather than final physical parameters.
The statistical yields, background contaminations, and the corresponding $\chi^2/ndf$ values are summarized in Table~\ref{tab:ml_fit_result}.
To ensure an unbiased performance evaluation, all metrics presented in Table~\ref{tab:ml_fit_result} were calculated using an independent test dataset entirely separated from the training and validation phases.

\begin{table}[htbp]
  \centering
  \caption{Comparison of signal statistics, background contamination, and fitting accuracy ($\chi^2/ndf$) in the cusp region ($2124$--$2139$ MeV/$c^2$). The $\chi^2/ndf$ values are presented as auxiliary indicators for relative comparison. Values are presented as the mean $\pm$ standard deviation from three independent iterations.}
  \label{tab:ml_fit_result}
  \begin{tabular}{lccc}
    \toprule
    Condition & \begin{tabular}[c]{@{}c@{}}Stats.\\(2050--2250 MeV/$c^2$)\end{tabular} & \begin{tabular}[c]{@{}c@{}}BG Contamination\\(2124--2139 MeV/$c^2$)\end{tabular} & \begin{tabular}[c]{@{}c@{}}$\chi^2 / \text{ndf}$\\(2124--2139 MeV/$c^2$)\end{tabular} \\
    \midrule
    Inclusive & 76,000 & $68.22 \pm 0.01$\% & --- \\
    $\mathrm{Mt}=3$ & 13,180 & $1.77 \pm 0.01$\% & $2.23 \pm 0.19$ \\
    $\mathrm{Mt}=2$ & 29,369 & $10.94 \pm 0.03$\% & $8.93 \pm 0.09$ \\
    $\mathrm{Mt}=2+\mathrm{ML}$ & 14,645 & $2.10 \pm 0.02$\% & $1.44 \pm 0.06$ \\
    \bottomrule
  \end{tabular}
\end{table}

The ML-based selection (Mt=2+ML) successfully secured 14,645 events, exceeding the statistics of the traditional Mt=3 condition, while achieving a comparable background contamination level ($2.10\%$).
More importantly, the relative fitting accuracy improved remarkably to $\chi^2/ndf = 1.44$, vastly outperforming both the unselected Mt=2 condition ($\approx 8.9$) and the Mt=3 condition ($\approx 2.2$).

\subsection{Verification of the Spectral Acceptance Bias}

\begin{figure}[htbp]
  \centering
  \includegraphics[width=1.0\linewidth]{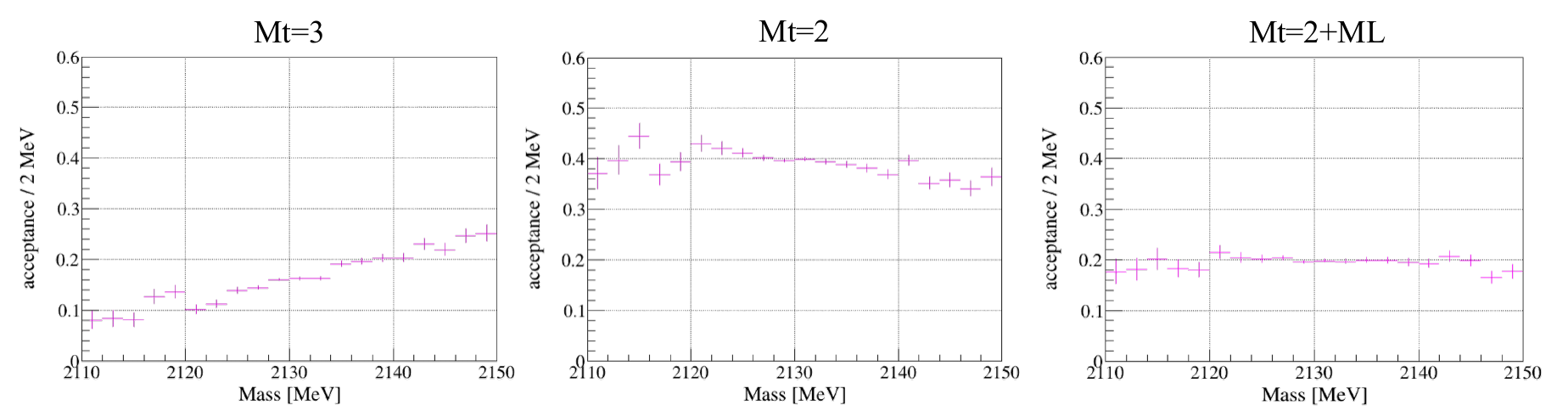}
  \caption{Mass dependence of the acceptance for different selection methods. Unlike the $\mathrm{Mt}=3$ condition, the $\mathrm{Mt}=2+\mathrm{ML}$ selection exhibits a completely flat response, avoiding spectral shape distortion.}
  \label{fig:mm_cuspfit}
\end{figure}

The significant improvement in the relative fitting accuracy ($\chi^2/ndf$) can be attributed to the preservation of the intrinsic spectral shape.
To verify this, we evaluated the mass dependence of the acceptance (the ratio of events surviving the selection), as shown in Fig.~\ref{fig:mm_cuspfit}.
While the Mt=3 condition exhibits a distinct right-upward slope (shape distortion) due to the kinematic efficiency drop at lower masses, the Mt=2+ML condition demonstrates an extremely flat acceptance across the threshold region.
The physical reasons behind this flatness and its implications for the experimental analysis are discussed in the following section.

\section{Discussion}\label{sec:discussion}

\subsection{Interpretability of the ML Model with SHAP}
\begin{figure}[htbp]
  \centering
  \includegraphics[width=1.0\linewidth]{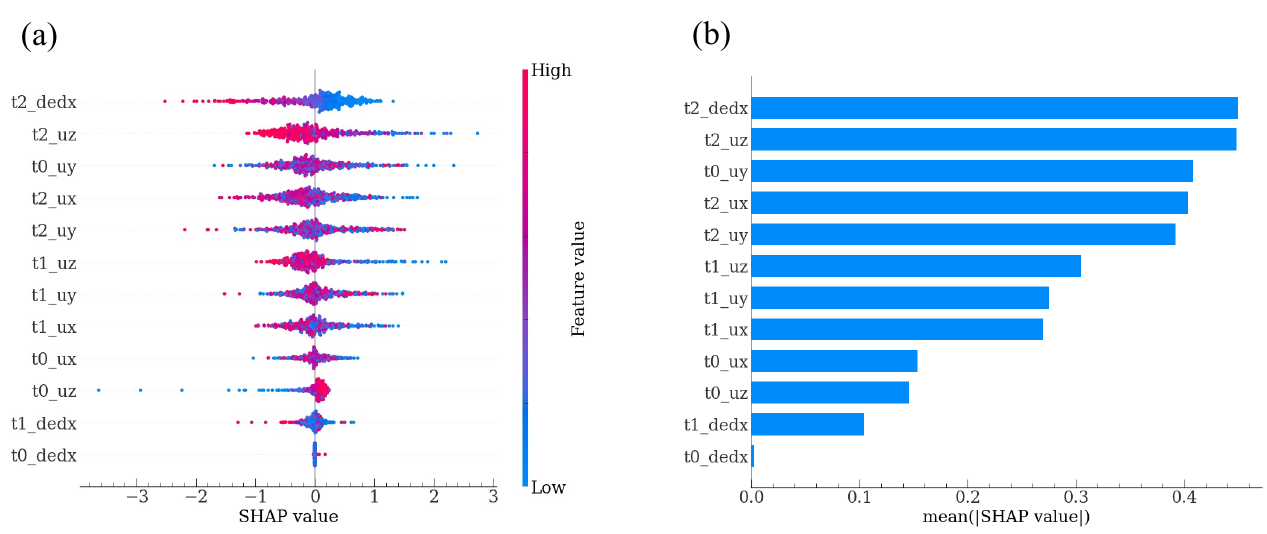}
  \caption{Feature importance and impact evaluated using SHAP. (a) Summary plot showing the distribution of SHAP values for each feature. The color represents the feature value (red for high, blue for low). (b) Bar plot ranking the features by their mean absolute SHAP values (overall contribution).}
  \label{fig:shap}
\end{figure}

Deep learning models are often regarded as ``black boxes,'' but understanding the basis of their classification is crucial in physics experiments to ensure the physical validity of the selection.
To interpret the model's decision-making process, we employed SHAP (SHapley Additive exPlanations) \cite{lundberg2017}, a game-theoretic approach to explain the output of machine learning models.
SHAP calculates the Shapley value, which quantifies the marginal contribution of each feature to the model's final prediction.
Mathematically, the Shapley value $\phi_i$ for a feature $i$ is defined as the weighted average of its marginal contributions over all possible combinations of features:
\begin{equation}
    \phi_i = \sum_{S \subseteq F \setminus \{i\}} \frac{|S|! (|F| - |S| - 1)!}{|F|!} \left[ f(S \cup \{i\}) - f(S) \right],
\end{equation}
where $F$ is the set of all input features, $S$ is a subset of features not containing $i$, and $f(S)$ is the model's prediction using only the subset $S$.
The term $f(S \cup \{i\}) - f(S)$ represents the change in the prediction (marginal contribution) when feature $i$ is introduced to the subset $S$.
In our binary classification context, a positive SHAP value ($\phi_i > 0$) indicates that the feature pushes the prediction toward the ``Signal'' class, while a negative value pushes it toward the ``Background'' class.
The overall feature importances and their impacts on the model's output are visualized in Fig.~\ref{fig:shap}.

As shown in the bar plot (Fig.~\ref{fig:shap}(b)), the most influential features for the model's decision were \texttt{t2\_dedx} (energy loss of Track 2, the decay proton candidate) and \texttt{t2\_uz} (the beam-axis direction cosine of Track 2).
By carefully examining the summary plot (Fig.~\ref{fig:shap}(a)), we can draw the following physical interpretations:

\begin{itemize}
    \item \textbf{Impact of \texttt{t2\_dedx}:} In regions where \texttt{t2\_dedx} is relatively small (blue points), the SHAP value is positive, contributing to a signal classification. This indicates that the model effectively identifies and rejects low-momentum (high $dE/dx$) protons originating from QF reactions, thereby efficiently extracting the signal proton components.
	\item \textbf{Impact of \texttt{t2\_uz}:} High values of \texttt{t2\_uz} (red points), corresponding to a small scattering angle $\theta$ in the forward direction, strongly contribute to a background classification (negative SHAP value). As discussed in Section~\ref{sec:multiplicity}, the QF $\Lambda$ is typically emitted forward with a large recoil momentum, causing its decay proton to be concentrated in the forward direction. The model accurately captures this kinematic distinction.
\end{itemize}

These observations perfectly align with the qualitative physical expectations, demonstrating that the neural network has learned physically valid criteria.
It is noteworthy that while simple one-dimensional cuts fail due to the extensive overlap of these variables in the parameter space, the neural network organically captures the multi-dimensional correlations among various components (including \texttt{t0\_uy}, \texttt{t2\_ux}, etc.) to achieve sophisticated classification.

\subsection{Implications for Systematic Error Evaluation}
The flat acceptance of the Mt=2+ML selection, observed in Fig.~\ref{fig:mm_cuspfit}, is a direct consequence of the input feature design analyzed above.
Crucially, neither the absolute momentum of the scattered $\pi^-$ nor the reconstructed missing mass itself was provided to the network.
Since the model relies solely on the multi-dimensional topological and kinematic correlations of the decay products ($dE/dx$ and direction cosines) to filter out backgrounds, it naturally avoids introducing a mass-dependent bias.
This characteristic is a powerful asset for real experimental data analysis.
The shape distortion inherent in the Mt=3 selection necessitates an acceptance correction based on Monte Carlo simulations, where any imperfection in the simulation translates directly into a systematic error in the final scattering length determination.
Conversely, by utilizing the unbiased Mt=2+ML event sample, we can cross-validate the acceptance corrections directly from the data.
Thus, the proposed ML approach not only doubles the total analyzable statistics by combining independent event samples (Mt=3 and Mt=2+ML), but also provides an indispensable tool for mitigating and evaluating systematic errors, thereby guaranteeing the reliability of the high-precision $\Sigma N$ cusp spectroscopy.

\section{Conclusion}\label{sec:conclusion}

The precise determination of the $\Sigma N$ scattering length is crucial for understanding the hyperon--nucleon interaction and its implications for the equation of state of neutron stars.
The J-PARC E90 experiment aims to achieve this through high-resolution missing-mass spectroscopy of the $\Sigma N$ cusp.
However, the experiment faces a critical challenge: suppressing the dominant QF background processes without compromising statistical power or distorting the intrinsic spectral shape.

In this study, we evaluated the conventional event selection based on the track multiplicity (Mt) in the HypTPC and developed a novel background suppression technique using machine learning.
Our Geant4-based simulations revealed a fundamental trade-off in the conventional methods:

\begin{itemize}
    \item The Mt=3 selection strongly suppresses backgrounds but suffers from severe statistical loss and introduces a mass-dependent acceptance bias that distorts the spectrum.
    \item The Mt=2 selection retains high statistics and a flat acceptance but leaves a substantial amount of QF background, making precise spectral fitting impossible.
\end{itemize}

To resolve this trade-off, we implemented a Neural Network (NN) to classify Mt=2 events.
By learning the multi-dimensional topological and kinematic correlations of the decay products ($dE/dx$ and direction cosines) and employing optimization techniques such as class weighting, early stopping using the F1 score, and hyperparameter tuning via Optuna, we constructed a highly robust classification model.

\begin{figure}[htbp]
  \centering
  \includegraphics[width=0.8\linewidth]{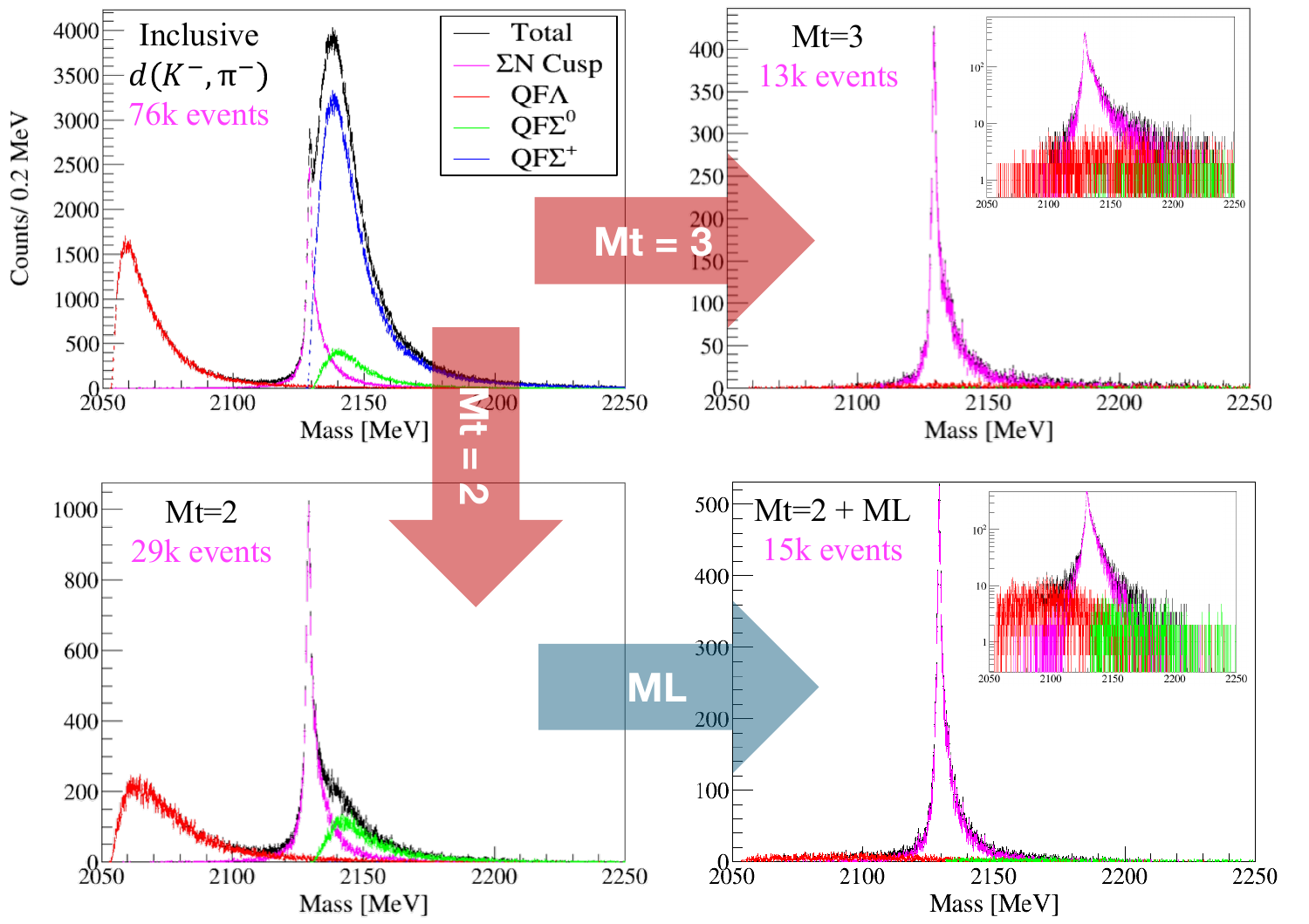}
  \caption{Comparison of the simulated missing-mass spectra. Top left: the inclusive measurement expected in the experiment. The other panels show the spectra after applying different background suppression methods. The ML-based selection ($\mathrm{Mt}=2+\mathrm{ML}$) dramatically suppresses the QF backgrounds while maintaining high statistics and an unbiased signal shape.}
  \label{fig:mmcomp}
\end{figure}

As summarized in Table~\ref{tab:method_comparison_summary} and visually compared in Fig.~\ref{fig:mmcomp}, the application of the developed ML selection (Mt=2+ML) successfully achieved the following milestones:
\begin{enumerate}
    \item \textbf{Simultaneous High S/N and High Statistics:} The background contamination in the critical cusp region was suppressed to $2.10\%$, rivaling the purity of the Mt=3 condition, while retaining 14,645 signal events, thereby recovering the lost statistical power.
    \item \textbf{Avoidance of Spectral Bias:} Because explicit mass or momentum information was excluded from the input features, the NN naturally avoided mass-dependent cuts, resulting in an exceptionally flat acceptance across the threshold region.
    \item \textbf{Improvement in Relative Fitting Accuracy:} As a result of the high purity and the unbiased spectral shape, the relative fitting accuracy improved drastically. In our uncorrected test fits, the ML-based selection yielded a significantly better reduced chi-square compared to the conventional methods, demonstrating superior preservation of the intrinsic spectral shape.
\end{enumerate}

\begin{table}[htbp]
  \centering
  \caption{Summary of performance comparisons for different event selection methods. The signal statistics are evaluated in the range of $2050$--$2250$ MeV/$c^2$, while the background contamination and $\chi^2/ndf$ are evaluated in the cusp peak region ($2124$--$2139$ MeV/$c^2$).}
  \label{tab:method_comparison_summary}
  \begin{tabular}{lcccc}
    \toprule
    Method & Stats. & BG Contamination & Acceptance Bias & $\chi^2 / ndf$ \\
    \midrule
    Inclusive & 76,000 & $68.22 \pm 0.01$\% & --- & --- \\
    $\mathrm{Mt}=3$ & 13,180 & $1.77 \pm 0.01$\% & Present & $2.23 \pm 0.19$ \\
    $\mathrm{Mt}=2$ & 29,369 & $10.94 \pm 0.03$\% & Absent & $8.93 \pm 0.09$ \\
    \rowcolor[gray]{0.9}
    $\mathrm{Mt}=2+\mathrm{ML}$ & 14,645 & $2.10 \pm 0.02$\% & Absent & $1.44 \pm 0.06$ \\
    \bottomrule
  \end{tabular}
\end{table}

Furthermore, the Mt=3 and Mt=2+ML conditions handle mutually independent event samples.
By combining both methods, we can essentially double the total analyzable statistics compared to the conventional approach.
More importantly, cross-referencing the results from the biased Mt=3 analysis with the unbiased Mt=2+ML analysis allows for a data-driven evaluation of systematic errors related to acceptance corrections.

In conclusion, the machine learning-based background suppression technique established in this study provides a highly balanced and powerful framework.
It significantly enhances both the statistical precision and the systematic reliability of the experimental results, paving the way for the definitive extraction of the $\Sigma N$ scattering length in the J-PARC E90 experiment.

Looking ahead to the application of this method to real experimental data, it is essential to address the potential risk of ``domain shift''---a degradation in model performance caused by small discrepancies between simulated and actual detector responses, such as $dE/dx$ resolution or track reconstruction efficiency in the HypTPC.
To guarantee the reliability of the ML selection in actual physics run, a robust, data-driven validation strategy using control samples will be essential.
For instance, high-purity control samples from well-known two-body kinematic reactions (e.g., $K^- p \to \Sigma^+ \pi^-$ and $K^- p \to \Sigma^- \pi^+$) can be extracted using the forward S-2S spectrometer, completely independent of the HypTPC tracking.
By comparing these actual, clean track data with our Geant4 simulations, detector responses can be evaluated.
Furthermore, feeding these real tracks into the ML model will allow us to directly quantify the classification efficiency and validate the decision boundaries on experimental data.
Alongside evaluating systematic uncertainties by varying the classification threshold, this data-driven validation strategy is expected to ensure that the proposed ML methodology remains robust and credible for the ultimate physics goal of the E90 experiment.

\section*{Acknowledgment}
This work was supported by JSPS KAKENHI Grant Number 25H01272.
The authors would like to express their sincere gratitude to the members of the J-PARC E90 Collaboration for their invaluable discussions and continuous support.
We also acknowledge the use of artificial intelligence tools for their assistance in coding and manuscript preparation.

\FloatBarrier

\vspace{0.2cm}
\noindent
\let\doi\relax


\end{document}